\documentclass[12pt]{article}
\usepackage{epsfig}
\newcommand{\z}{&&\hspace*{-1cm}}

\newcommand{\bea}{\begin{eqnarray}}
\newcommand{\eea}{\end{eqnarray}}
\newcommand{\be}{\begin{equation}}
\newcommand{\ee}{\end{equation}}

\title{Gluon density in the rescaling model
}

\author{N.A.~Abdulov$^{1}$, A.V.~Kotikov$^{2}$, A.V.~Lipatov$^{1,2}$}

\begin{document}

\maketitle

\begin{center}
{\it $^{1}$Skobeltsyn Institute of Nuclear Physics, Lomonosov Moscow State University, 119991, Moscow, Russia}\\
{\it $^{2}$Joint Institute for Nuclear Research, 141980, Dubna, Moscow region, Russia}

\end{center}

\vspace{0.5cm}

\begin{center}

{\bf Abstract }
       
\end{center}

\indent
The behavior of the gluon density in nuclei is investigated in the framework of the
rescaling model.


\vspace{1.0cm}

\noindent{\it Keywords:}
Deep inelastic scattering; parton densities; EMC effect.

\vspace{1.0cm}

\section{Introduction}

The study of deep inelastic scattering (DIS) of leptons on nuclei reveals the appearance of a significant nuclear effect, which excludes the naive idea of
the nucleus as a system of quasi-free nucleons (see for an overview, for example,~\cite{Arneodo:1992wf,Rith:2014tma}).
This effect was first discovered by the European Muon Collaboration~\cite{Aubert:1983xm} in the domain of valence quark dominance, so it was named as the EMC effect.

Nowadays there are two main approaches to study the EMC effect. In the first one, which is currently more common, nuclear parton distribution functions (nPDFs) are
extracted from global fits (see a recent review~\cite{Paakkinen:2022qxn} and references therein) to nuclear data using empirical parameterization of their
normalizations and the numerical solution of the Dokshitzer-Gribov-Lipatov-Altarelli-Parisi (DGLAP)~\cite{DGLAP} equations.
The second strategy is based on some models of PDFs (see various models, e.g., in the original papers \cite{Kulagin:2004ie,Jaffe:1983zw,Close:1983tn}
and in a review~\cite{Kulagin:2016fzf}).

Here we will follow the scaling model  \cite{Jaffe:1983zw} based on the  assumption \cite{Close:1983tn} that the effective size of confinement of gluons and quarks
in a nucleus is larger than in a free nucleon. 
Within perturbative QCD, it was found \cite{Jaffe:1983zw,Close:1983tn} that this confinement scaling predicts that nPDF and regular (nucleon) PDFs can be related
by simple scaling of the $Q^2$ argument.
Thus, we can say that the rescaling model demonstrates features inherent in both approaches: within its framework, there are certain relationships between conventional
and nuclear PDFs that arise as a result of a shift in values of the kinematic variable $Q^2$ and, at the same time, both densities obey the DGLAP equations.

Initially, the rescaling model was created for the valence quark dominance region $0.2 \leq x \leq 0.8$.
Recently in Refs.~\cite{Kotikov:2017mhk,Kotikov:2018ass}, its applicability was extended to a range of small values of $x$, where certain shadowing and antishadowing
effects were found for the sea quark and gluon densities.

The purpose of this short paper is to apply the rescaling model to the recently published PDF parametrizations \cite{Abdulov:2022itv}
and show the nuclear modification of the gluon density in a wider range of $x$.

\section{Sea quark and gluon densities}

Using the $Q^2$-evolution of PDFs for large and small values of $x$ (see \cite{Gross,LoYn} and \cite{Rujula,Q2evo} respectively, as well as the review in \cite{Kotikov2007}),
the new type of PDF parametrizations was constructed in Ref. \cite{Abdulov:2022itv}.
\footnote{In a sense Ref. \cite{Abdulov:2022itv} is a continuation of previous studies \cite{Illarionov:2010gy} carried out for valence quarks.}

The sea quark and gluon parts can be represented as combinations of $\pm$ terms:
\bea
\z f_a(x,Q^2) = \sum_{\pm}\, f_{a,\pm}(x,Q^2),~~ a=q,g,~~ \nonumber \\
\z f_{q,-}(x,Q^2)=
\biggl[A_{q}e^{- d_{-} s} (1-x)^{m_{q,-}} +   \frac{B_{-}(s)\, x}{\Gamma(1+\nu_{-}(s))}+ D_{-}(s)x (1 -x)  \biggr]\,  (1-x)^{\nu_{-}(s)},\nonumber \\
\z f_{g,-}(x,Q^2)=
A_{g}^{-}e^{- d_{-} s}
\,  (1-x)^{\nu_{-}(s)+m_{g,-}+1}, \nonumber \\
\z f_{g,+}(x,Q^2)= 
\biggl[A_{g}^+ \overline{I}_0(\sigma)e^{-\overline d_{+} s} (1-x)^{m_{g,+}} + \frac{B_{+}(s)\, x}{\Gamma(1+\nu_{+}(s))} + D_{+}(s)x (1 -x)  \biggr]\, (1-x)^{\nu_{+}(s)}, \nonumber \\
\z f_{q,+}(x,Q^2)= 
A_{q}^+ \tilde{I}_1(\sigma)e^{-\overline d_{+} s}
  (1-x)^{\nu_{+}(s)+m_{q,+}+1}
  \,.\label{q+l}
\eea
where
\be
\nu_{\pm}(s)=\nu_{\pm}(0)+r_{\pm}s,~~B_{\pm}(s)=B_{\pm}(0) e^{p_{\pm}s},~~p_{\pm}=r_{\pm}\bigl(\gamma_{\rm E}+\hat{c}_{\pm}\big),
\label{nupm}
\ee
with $\gamma_{\rm E}$ is Euler's constant,
$f$ is a number of active quarks, $\beta_0=11-(2/3)f$ is the leading order of QCD $\beta$-function and
\be
r_{+}=\frac{12}{\beta_0},~~r_{-}=\frac{16}{3\beta_0},~~\hat{c}_{+}=-\frac{\beta_0}{12},~~\hat{c}_{-}=-\frac{3}{4}\,.
\label{rcpm}
\ee

Here $I_{\nu}$ ($\nu=0,1$)
are the modified Bessel functions
with
\be
\sigma = 2\sqrt{\left|\hat{d}_+\right| s
  \ln \left( \frac{1}{x} \right)}  \, ,~~ \rho=\frac{\sigma}{2\ln(1/x)},~~
  \hat{d}_+ = - \frac{12}{\beta_0},~~
\overline d_{+} = 1 + \frac{20f}{27\beta_0},~~
d_{-} = \frac{16f}{27\beta_0} \, .
\label{intro:1a}
\ee
The factors $Q_0^2$, $A_a$, $B_{\pm}(0)$ and $\nu_{\pm}(0)$ are free parameters obtained in~\cite{Abdulov:2022itv}.
The $Q^2$ dependence of the subasymptotic terms $\sim D_{\pm}(s)$ is taken from the momentum conservation law.
In \cite{Abdulov:2022itv} $m_{q,-}=m_{g,+}=2$ and $m_{q,+}=m_{g,-}=1$. In this case, the small $x$ asymptotics are suppressed at large $x$ compared to
the subasymptotic terms $\sim D_{\pm}(s)$.
Moreover, the small $x$ asymptotics contain the same powers of the factor $(1-x)$ for quarks and gluons.

\section{Rescaling model}

In the rescaling model~\cite{Jaffe:1983zw}, the DIS structure function $F_2$ and, consequently, the valence part of quark densities are modified in the case of
a $A$ nucleus at intermediate and large values of $x$ $( 0.2 \leq x \leq 0.8)$ as follows
\begin{equation}
  f_{V}^A(x,Q^2) =
  f_{V}(x,Q^2_{A,V}),
  \label{va.1}
\end{equation}
where the new scale $Q^2_{A,V}$ is related to $Q^2$ by \cite{Kotikov:2017mhk}
\begin{equation}
s^A_V \equiv \ln \left(\frac{\ln\left(Q^2_{A,V}/\Lambda^2\right)}{\ln\left(Q^2_{0}/\Lambda^2\right)}\right)
= s +\ln\Bigl(1+\delta^A_V\Bigr) \approx s +\delta^A_V,~~~
\label{sA}
\end{equation}
i.e. the kernel modification of the main variable $s$ depends on the
$Q^2$-independent parameter $\delta^A_V$ having small values (see Tables 2 and 3 in~\cite{Kotikov:2017mhk}).

\subsection{Rescaling model at low $x$}

In~\cite{Kotikov:2017mhk} the PDF asymptotics for small $x$ shown in (\ref{q+l}) were applied to the small $x$ region of the EMC effect using the simple fact
that the rise of sea quark and gluons densities increases with increasing values of $Q^2$.
Thus, in the case of nuclei, the PDF evolution scale is less than $Q^2$ and this can directly reproduce the shadowing effect observed in global fits.
Since there are two components for each parton density (see Eq.~(\ref{q+l})), we have two free parameters $Q^2_{A,\pm}$ that can be determined from the analysis of
experimental data for the EMC effect at low $x$ values.

Note that it is usually convenient to study the following ratio
\begin{equation}
R^{AD}_{g}(x,Q^2) = \frac{f^A_g(x,Q^2)}{f^D_g(x,Q^2)}\,.
\label{AD}
\end{equation}

Taking advantage of the fact that the nuclear effect in the deuteron is very small (see
discussions in~\cite{Kulagin:2016fzf})
\footnote{Study of nuclear effects in the deuteron can be found in a recent article~\cite{AKP}, which also contains summaries of preliminary studies.}
:  $f^D_a(x,Q^2) \approx f_a(x,Q^2)$, we can assume that
\be
  f^{A}_a(x,Q^2)
~=~
f_a^{A,+}(x,Q^2) + f_a^{A,-}(x,Q^2),~~
f^{A,\pm}_a(x,Q^2) =
f^{\pm}_a(x,Q^2_{AD,\pm}) \, .
\label{AD1}
\ee

The expressions for $f^{\pm}_a(x,Q^2)$ are given in Eq.~(\ref{q+l}) and the corresponding values of $s^{AD}_{\pm} $ turned out to be
\be
s^{AD}_{\pm} \equiv \ln \left(\frac{\ln\left(Q^2_{AD,\pm}/\Lambda^2\right)}{\ln\left(Q^2_{0}/\Lambda^2\right)}\right)
  = s +\ln\Bigl(1+\delta^{AD}_{\pm}\Bigr)\,,
\label{AD2}
\ee
where the results for $\delta^{AD}_{\pm}$ can be found in Ref. ~\cite{Kotikov:2017mhk}.

\section{Results}

\begin{figure}
\centering
\vskip 0.5cm
\includegraphics[width=17.0cm]{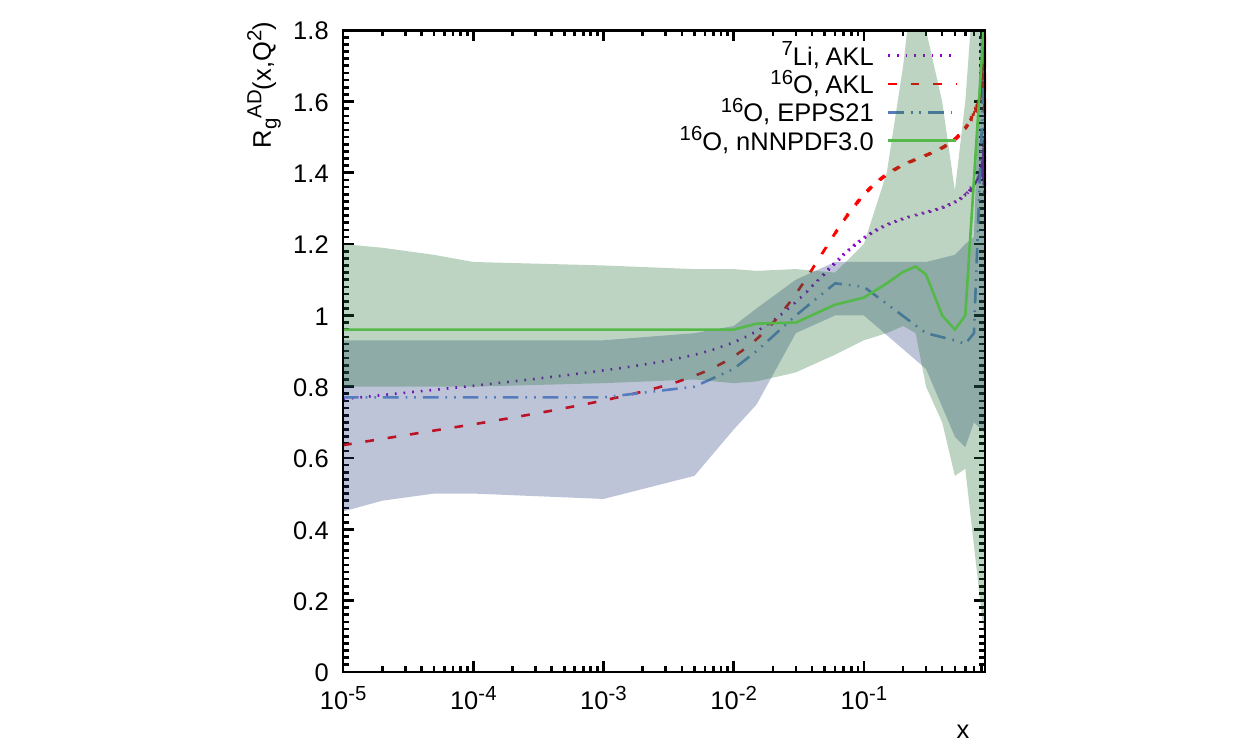}
\vskip -0.3cm
\caption{$x$ dependence of $R^{AD}_{g}(x,Q^2)$
at $Q^2$=10 GeV$^2$. The purple dotted and red dashed lines are for the results obtained
in the present paper for ${}^7$Li and ${}^{16}$O, respectively. The blue dash-dotted and green lines and bands
are borrowed from the recent review~\cite{Paakkinen:2022qxn} (see Fig. 1 in \cite{Paakkinen:2022qxn}).
}
\end{figure}

The results obtained for $R^{AD}_{g}(x,Q^2)$ are shown in Fig. 1. As one can see, our results are very close to the results obtained in \cite{Kotikov:2017mhk}
at $x \leq 10^{-2}$, since the parameters $\delta_{\pm}^{AD}$ are taken from that paper. Moreover, in the low $x$ range, our results are also close to the recent
results obtained by fitting experimental data and shown in the review \cite{Paakkinen:2022qxn}.

However, we see a discrepancy between our results and the results of \cite{Paakkinen:2022qxn} at large values of $x$: $x \geq 0.1$. This must be due to
the use of the $\delta_{\pm}^{AD}$ parameters, which were obtained in \cite{Kotikov:2017mhk} only during the study of the experimental data \cite{Arneodo:1995cs,Adams:1995is} obtained at small $x$ values.
In addition, since the \cite{Arneodo:1995cs,Adams:1995is} data were obtained at very low $Q^2$ values, the quality of the fit performed in \cite{Kotikov:2017mhk} is not
very good. For example, the fitting results obtained using two different infrared modifications of the strong coupling constant: analytical and ``frozen'',
(see Ref.\cite{Kotikov:2004uf} and discussions therein) are very different, which is very unusual.
We hope to study this phenomenon in our future publications, where we plan to take into account the new parametrizations (\ref{q+l}) that are valid over the entire $x$ range
and experimental data \cite{Aubert:1983xm,Arneodo:1995cs,Adams:1995is} from different ranges of the Bjorken variable $x$.

In addition, we plan to study nuclear modifications of unintegrated PDFs \cite{Kotikov:2019kci,Abdulov:2022itv}, which are now becoming very popular
(see \cite{Abdulov:2021ivr} and references and discussions therein).\\


Researches described in Section~2 were 
supported by the Russian Science Foundation under grant 22-22-00119.
Studies described and performed in Sections 3 and 4 were supported by the Russian Science Foundation under grant 22-22-00387.

\end{document}